\documentclass[twocolumn,superscriptaddress]{revtex4-2}
\usepackage[english]{babel}
\usepackage{amsmath}
\usepackage{amssymb}
\usepackage{stmaryrd}   
\usepackage[retain-unity-mantissa=false]{siunitx}
\usepackage{graphicx}
\usepackage{xcolor}
\usepackage[caption=false]{subfig}
\usepackage{placeins}
\usepackage[colorlinks=true, linkcolor=blue, citecolor=blue, filecolor=blue, urlcolor=blue,pdfproducer={.},pdfcreator={.}]{hyperref}
\usepackage{cleveref}
\usepackage{blindtext}
\usepackage{comment}
\usepackage{tikz}
\usepackage{verbatim}
\usepackage{ulem}
\usepackage{physics}
\usepackage[commandnameprefix=ifneeded, commentmarkup=uwave]{changes}

\definechangesauthor[name=Christopher, color=orange]{C}
\definechangesauthor[name=Anulekha, color=cyan]{ade}

\newcommand{\mutm}{$\mu T$ model }

\newcommand{\nm}{\nano\meter}

\begin{document}

\title{Dynamic interfacial effects in ultrathin ferromagnetic bilayers}
\author{Anulekha \surname{De}}
\email{ade@rptu.de}
\affiliation{Fachbereich Physik and Landesforschungszentrum OPTIMAS, Rheinland-Pfälzische Technische Universität Kaiserslautern-Landau, 67663 Kaiserslautern, Germany}
\author{Christopher \surname{Seibel}}
\email{cseibel@rptu.de}
\affiliation{Fachbereich Physik and Landesforschungszentrum OPTIMAS, Rheinland-Pfälzische Technische Universität Kaiserslautern-Landau, 67663 Kaiserslautern, Germany}
\author{Sanjay \surname{Ashok}}
\affiliation{Fachbereich Physik and Landesforschungszentrum OPTIMAS, Rheinland-Pfälzische Technische Universität Kaiserslautern-Landau, 67663 Kaiserslautern, Germany}
\author{Paul \surname{Herrgen}}
\affiliation{Fachbereich Physik and Landesforschungszentrum OPTIMAS, Rheinland-Pfälzische Technische Universität Kaiserslautern-Landau, 67663 Kaiserslautern, Germany}
\author{Akira \surname{Lentfert}}
\affiliation{Fachbereich Physik and Landesforschungszentrum OPTIMAS, Rheinland-Pfälzische Technische Universität Kaiserslautern-Landau, 67663 Kaiserslautern, Germany}
\author{Laura \surname{Scheuer}}
\affiliation{Fachbereich Physik and Landesforschungszentrum OPTIMAS, Rheinland-Pfälzische Technische Universität Kaiserslautern-Landau, 67663 Kaiserslautern, Germany}
\affiliation{University of Delaware, Newark, DE 19716, USA}
\author{Georg \surname{von Freymann}}
\affiliation{Fachbereich Physik and Landesforschungszentrum OPTIMAS, Rheinland-Pfälzische Technische Universität Kaiserslautern-Landau, 67663 Kaiserslautern, Germany}
\affiliation{Fraunhofer Institute for Industrial Mathematics ITWM, 67663 Kaiserslautern, Germany}
\author{Philipp \surname{Pirro}}
\affiliation{Fachbereich Physik and Landesforschungszentrum OPTIMAS, Rheinland-Pfälzische Technische Universität Kaiserslautern-Landau, 67663 Kaiserslautern, Germany}
\author{Baerbel \surname{Rethfeld}}
\affiliation{Fachbereich Physik and Landesforschungszentrum OPTIMAS, Rheinland-Pfälzische Technische Universität Kaiserslautern-Landau, 67663 Kaiserslautern, Germany}
\author{Martin \surname{Aeschlimann}}
\affiliation{Fachbereich Physik and Landesforschungszentrum OPTIMAS, Rheinland-Pfälzische Technische Universität Kaiserslautern-Landau, 67663 Kaiserslautern, Germany}

\begin{abstract}
We investigate the magnetization dynamics of an ultrathin Co (1.5 nm) /Py (1.5 nm) bilayer system from femtosecond (fs) to nanosecond (ns) timescales. Magnetization dynamics in the fs timescales is characterized as a highly non-equilibrium regime due to an ultrafast reduction of magnetization by laser excitation. On the other hand, the dynamics in the ns timescales is characterized as a close-to-equilibrium regime involving the excitation of coherent magnons. We demonstrate that the interfacial interaction between the Co and Py layers in these two non-equilibrium regimes across the timescales is dynamic and simultaneously influences the magnetization loss in the fs timescales and the magnon dynamics in the ns timescales. On ultrafast (fs) timescales, comparison between time-resolved magneto-optical Kerr effect (TR-MOKE) measurements and temperature-based $\mu T$ model simulations reveals that the bilayer exhibits demagnetization dynamics intermediate between those of its individual layers. When driven far from equilibrium by ultrashort laser pulse excitation, the magnetization dynamics of the individual Co and Py layers appear to remain decoupled and evolve independently in the initial stages of the ultrafast response. On the other hand, in the ns regime, the two individual layers of the bilayer precess together at the same frequency in a coupled manner as one effective single layer.  Furthermore, by correlating the ultrafast demagnetization to precessional damping we attempt to bridge the two non-equilibrium regimes across fs to ns timescales. These results improve our understanding of magnetization dynamics across timescales in ultrathin exchanged-coupled ferromagnetic bilayers and provide valuable insights for the design of high-frequency and energy efficient spintronic device concepts.

\end{abstract}

\date{\today}

\maketitle

\section{Introduction}

Ultrafast magnetometry is a cutting edge methodology to study light–matter interactions across timescales ranging from few attoseconds to several nanoseconds in various material systems. This methodology utilizes light pulses with a typical temporal width of a few femtoseconds that can excite as well as detect the magnetization dynamics in magnetic materials with a high spatiotemporal resolution.  The demonstration of ultrafast magnetization dynamics in  femtosecond laser irradiated thin nickel films by Beaurepaire et al.~\cite{Beaurepaire1996}  led to the inception of femtomagnetism. The phenomena of laser-induced magnetization dynamics range from the highly non-equilibrium regime on femtosecond timescales \cite{Koopmans2000, Bigot2009, Krauss2009, Koopmans2010, Malinowski2008, Battiato2010} to the close to equilibrium regime on picosecond to nanosecond timescales \cite{vanKampen2002, Koopmans2005unifying}.  Ultrafast magnetometry is typically used to study femtomagnetism in homogeneous films \cite{Beaurepaire1996, suchetana2023} as well as nano- and heterostructures \cite{anulekha_2021_PRB, Malinowski2008, anulekha_acsami_2022}.  Nevertheless, a clear understanding of the microscopic mechanisms that underlie this cross-timescale phenomenon as well as the methodologies to investigate them in various material systems are incomplete. A wide range of research has been performed on various ferromagnetic systems to understand the mechanisms that underpin femtomagnetism in sub-picosecond timescales.  In this context, two primary microscopic mechanisms are proposed to underlie ultrafast demagnetization: local spin-flip scattering (SFS) \cite{Koopmans2010, Zhang2009, Krauss2009, Bigot2009} and non-local spin-transport (ST) \cite{Malinowski2008, Battiato2010, Ashok2022} processes. The local SFS mechanisms encompass Elliott–Yafet (EY)-type electron-phonon scattering \cite{Koopmans2010}, electron-magnon interactions \cite{Zhang2009}, electron-electron Coulomb scattering \cite{Krauss2009}, and relativistic spin-flip events \cite{Bigot2009}. Regarding non-local spin transport, superdiffusive spin transport has been identified as the predominant mechanism contributing to the rapid loss of magnetization \cite{Malinowski2008, Battiato2010}.
It has also been discussed that ultrafast demagnetization is possibly a cooperative effect of the laser field and the spin-orbit coupling (SOC) of the magnetic material \cite{Zhang2000}. On the longer timescales (picosecond to nanosecond) the generation of coherent magnons are considered to play a strong role.  The signature of magnon generation can be observed as precessional dynamics on the nanosecond timescale, which represents a close-to-equilibrium regime \cite{vanKampen2002, Koopmans2005unifying}. This precession of magnetization is governed by the anisotropic magnetic properties, precessional frequencies, damping constants of magnetic materials and other material specific parameters \cite{vanKampen2002, Koopmans2005unifying, walowski2008, vomir2005}.

\par
The microscopic mechanisms underlying the ultrafast quenching of magnetization (in highly non-equilibrium region) and their relation to relaxation of magnetization through precession (in close-to-equilibrium region) remains unexplored. A simultaneous investigation of both the ultrafast demagnetization time and damping of the coherent magnons, along with the correlation between them, could contribute to study the transition from a highly non-equilibrium to a close-to-equilibrium region \cite{Dusabirane2023}. 
However, despite the considerable insights \cite{Koopmans2005unifying, Fahnle2010, suchetana2023}, the relationship between the two parameters is yet to be unambiguously clarified. 
A direct proportionality between demagnetization time and damping would hint at, for instance, local spin–flip scattering being the primary contributor to demagnetization \cite{Fahnle2010, suchetana2023}. An inverse proportionality, on the other hand, would support the non-local spin-transport mechanism as the dominant microscopic mechanism \cite{Koopmans2005unifying}.
\par
From a materials perspective, the investigation of this cross-timescale phenomenon has so far predominantly focused on single layer ferromagnet (FM), FM/nonmagnet (NM) or  FM/antiferromagnet (AFM) bilayers. However, only few works investigate to obtain a unified picture of ultrafast demagnetization and precession in ultrathin FM/FM bilayers\cite{Fahnle2017PRB}. Cobalt/Permalloy (Co/$\mathrm{Ni}_{80}\mathrm{Fe}_{20}$ or Co/Py) bilayers are one of the commonly studied FM/FM exchange spring material systems\cite{kneller1991ieeetrans, fullerton1998prb}. The exchange coupling between Co and Py is useful for studying buried magnetic interfaces and exchange effects in conducting structures. Understanding the interfacial effects in such exchange coupled bilayers is crucial to finely tailor their properties.
A number of works have focused on the magnetic properties of Co/Py bilayers, which have discussed magnon dynamics in such systems \cite{Crew2005, Kennewell_JAP_2010, DengJMMM2016, Rodríguez2006, arabinda2014, Salaheldeen2020, Feggeler2022}. Recently, Zhang et al. reported a direct relationship between demagnetization time and damping in Co/Ni bilayers \cite{Fahnle2017PRB}. Nevertheless, the existing literature lacks a deeper insight into the timescale-dependent dynamic interfacial effect, as well as a unified picture of ultrafast demagnetization and precessional damping in ultrathin, exchange coupled FM/FM bilayers.
\par

In our work, we report a comparative investigation of the magnetization dynamics of a Co/Py ultrathin bilayer system from femtosecond (fs) to nanosecond (ns) timescales using fs amplified laser pulses in a time-resolved magneto-optical Kerr effect (TR-MOKE) methodology. On ultrashort (fs) timescales, by relating experimental demagnetization to the simulations of magnetization using the temperature-based $\mu T$ model, we find that the demagnetization profiles of the bilayer is intermediate to that of the individual layers and that the two layers behave nearly independently in a decoupled way when the entire system is in the highly non-equilibrium state caused by ultrashort laser pulse excitation. However, on longer (ns) timescales, the experimentally measured precession dynamics of the ultrathin bilayer reveals a single precessional mode, indicating that the two layers precess together with a same frequency in a coupled manner, when the system is in close to equilibrium well after the laser irradiation is over. Finally, we study the correlation between ultrafast demagnetization and precessional (magnon) damping in the ultrathin bilayer. This comparison elucidates the dominant microscopic processes governing the transition of the system dynamics from highly non-equilibrium to the close-to-equilibrium regime after ultrafast laser pulse excitation. 
Our results show a direct proportionality between demagnetization time and magnon damping, which hints the predominance of local Elliott-Yafet (EY) processes in ultrathin FM/FM bilayers. 
\par
This paper is structured as follows. Section II describes the material system and its fabrication as well the theoretical and experimental methodologies of investigation. In section III, subsection A studies the magnetization dynamics in the highly non-equilibrium regime,  subsection B studies precession dynamics in close-to-equilibrium regime and subsection C provides an attempt to bridge the two non-equilibrium  regimes in a cross-timescale approach. Finally, section IV summarizes and concludes.

\begin{figure}[h]
    \centering
	\includegraphics[width=3.5in]{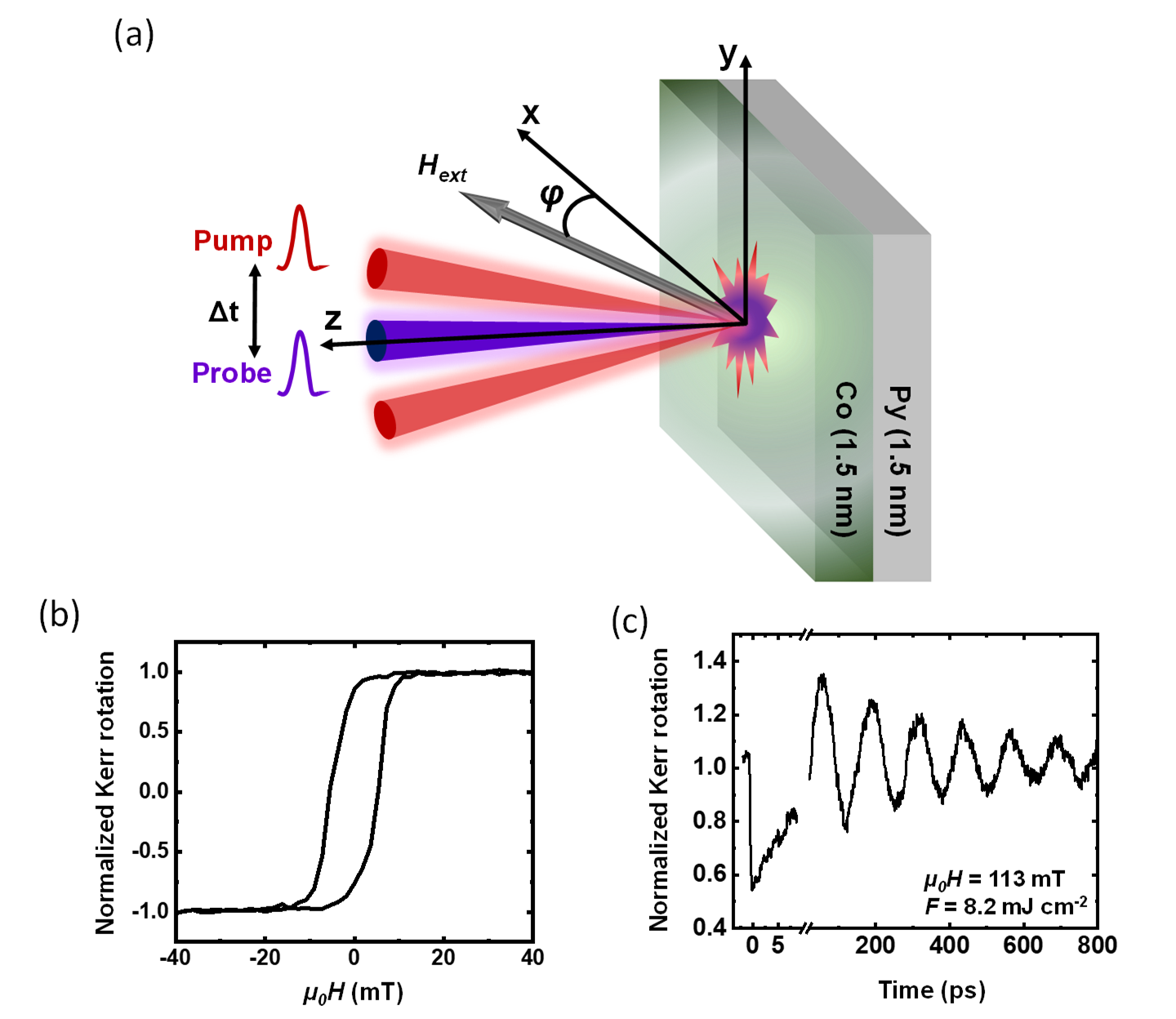}
	\caption{(a) Schematic of the pump-probe experiment (b)Hysteresis loop measured by static MOKE (c) representative time-resolved data measured at $\mu_{{0}}H$ = 113 mT and $F$ = 8.2 mJ cm$^{-2}$}
	\label{fig1}
\end{figure}

\section{materials and methods}

A Co/Py bilayer film is epitaxially grown on an MgO substrate by molecular beam epitaxy (MBE) in an ultrahigh vacuum chamber. The thickness of the Co and Py layers is 1.5 nm each. The sample is capped with \SI{3}{\nano\meter} $\mathrm{Al}_{2}\mathrm{O}_{3}$ and \SI{3}{\nano\meter} of Au to protect it from environmental degradation, oxidation and laser ablation during the pump-probe experiment. The sample structure and the experimental geometry is schematically shown in Fig.~\ref{fig1}(a). Magnetization dynamics from fs to ns timescales (i.e., from highly non-equilibrium to close to equilibrium regime) is measured using  time-resolved magneto optical Kerr effect (TR-MOKE) setup based on a two-color, non-collinear, all-optical pump-probe technique. Further details of the measurement technique is provided in the Supplemental Materials \cite{Supplement}. 
In order to calibrate the material properties in the equilibrium regime we study the hysteresis characteristics of the bilayer in the absence of excitation pulse. The in-plane hysteresis loop of the bilayer as shown in Fig.~\ref{fig1}(b) is measured by the same MOKE set up by blocking the pump beam. The rectangular shaped hysteresis loop indicates that in static case, the bilayer system behaves as a coupled system with one effective magnetic medium across the total thickness.
\par
The ultrafast magnetization dynamics of the heterostructure in the highly non-equilibrium regime (typically spanning sub-picosecond timescales) caused by excitation by a femtosecond laser pulse is modeled using the temperature-based \mutm for bilayers~\cite{Seibel2022,Mueller2014PRB}. 
This model traces the temporal evolution of temperatures and chemical potentials for spin-up and spin-down electrons separately for individual layers. The non-equilibrium in the spin-dependent temperatures and chemical potentials caused by laser excitation drives the initial magnetization dynamics. The internal coupling parameters for chemical potential and temperature coupling in the individual materials are obtained from previous experimental measurements for Co~\cite{Zusin2018} and Py~\cite{Anulekha2024PRB}, respectively. At the interface, spin-resolved particle and energy transfer processes are explicitly taken into account.  The spin-resolved electronic temperatures and chemical potentials determine spin-resolved Fermi distributions from which we determine the respective particle numbers of both the spin-up and spin-down reservoirs and thus the magnetization. The particle densities were obtained using the densities of states (DOS) for Co~\cite{Batallan1975} and Py~\cite{He2019}.
Both layers are homogeneously excited by the pump laser pulse. We calculate the absorption of the individual layers with the transfer matrix method from the refractive indices for the experimental wavelength of \SI{800}{\nano\meter}. Further details of the model can be found in Ref.~\cite{Seibel2022}. All parameters of the calculation are listed in the Supplemental Materials \cite{Supplement}.

\begin{figure}[t]
    \centering
	\includegraphics[width=3.4 in]{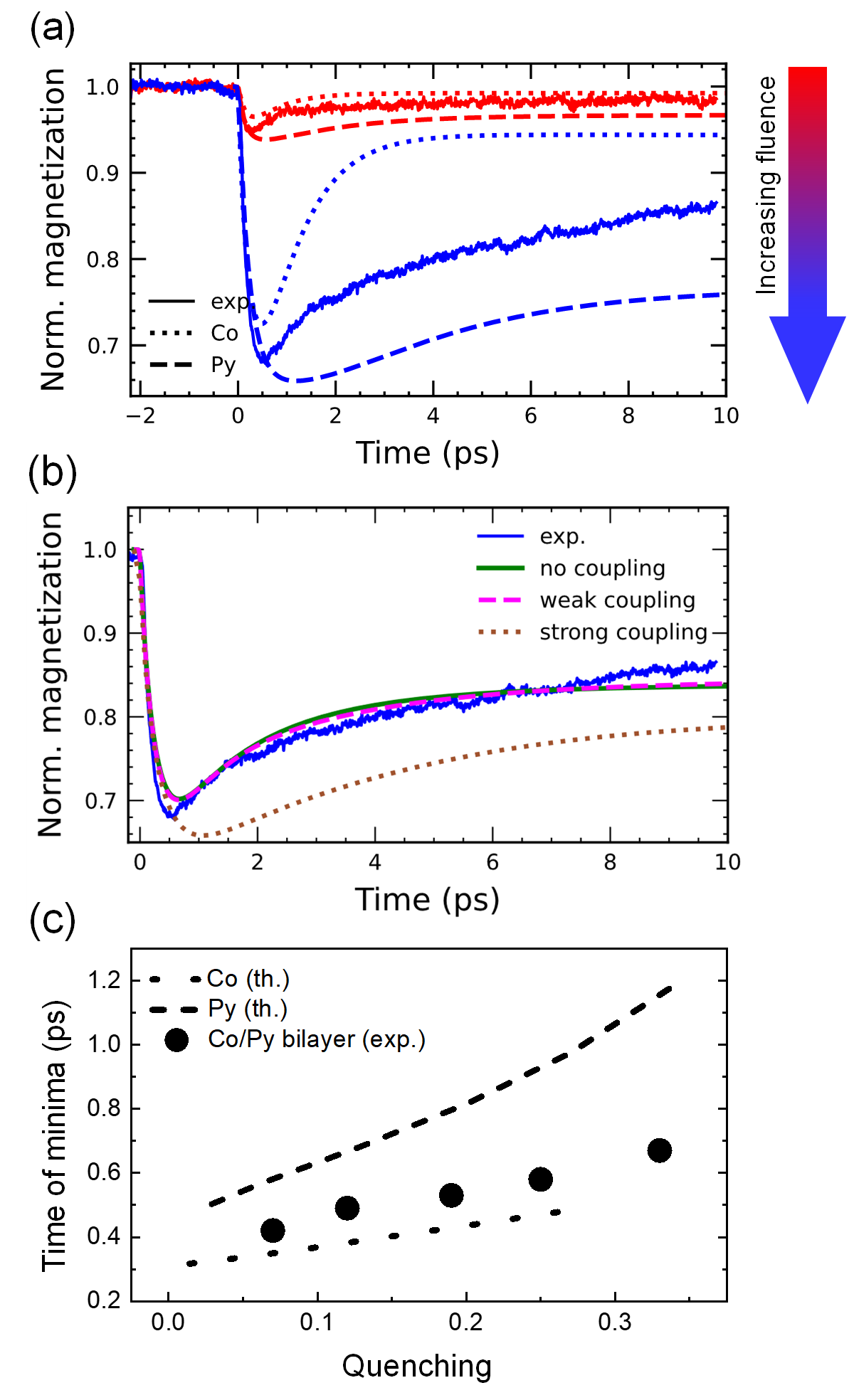}
	\caption{Comparison of the experimental  demagnetization dynamics of the Co/Py bilayer with theoretical $\mu T$ model. (a) The magnetization curves of the bilayer for one selected low (red) and one high (blue) fluence is in between the contributions from the individual layers of the bilayer. (b) Combined theoretical magnetization for various coupling strengths in comparison to the experimental data. (c) The time of minima vs. quenching determined by the fluence for the experimental bilayer and for the individual layers from theory. The results indicate decoupled dynamics at ultrafast timescales (see text).}
	\label{fig:ultrafast_magnetization}
\end{figure}

\section{Results and Discussion}
\subsection{Dynamics in the highly non-equilibrium regime}

\begin{figure}[t]
    \centering
	\includegraphics[width=3.5 in]{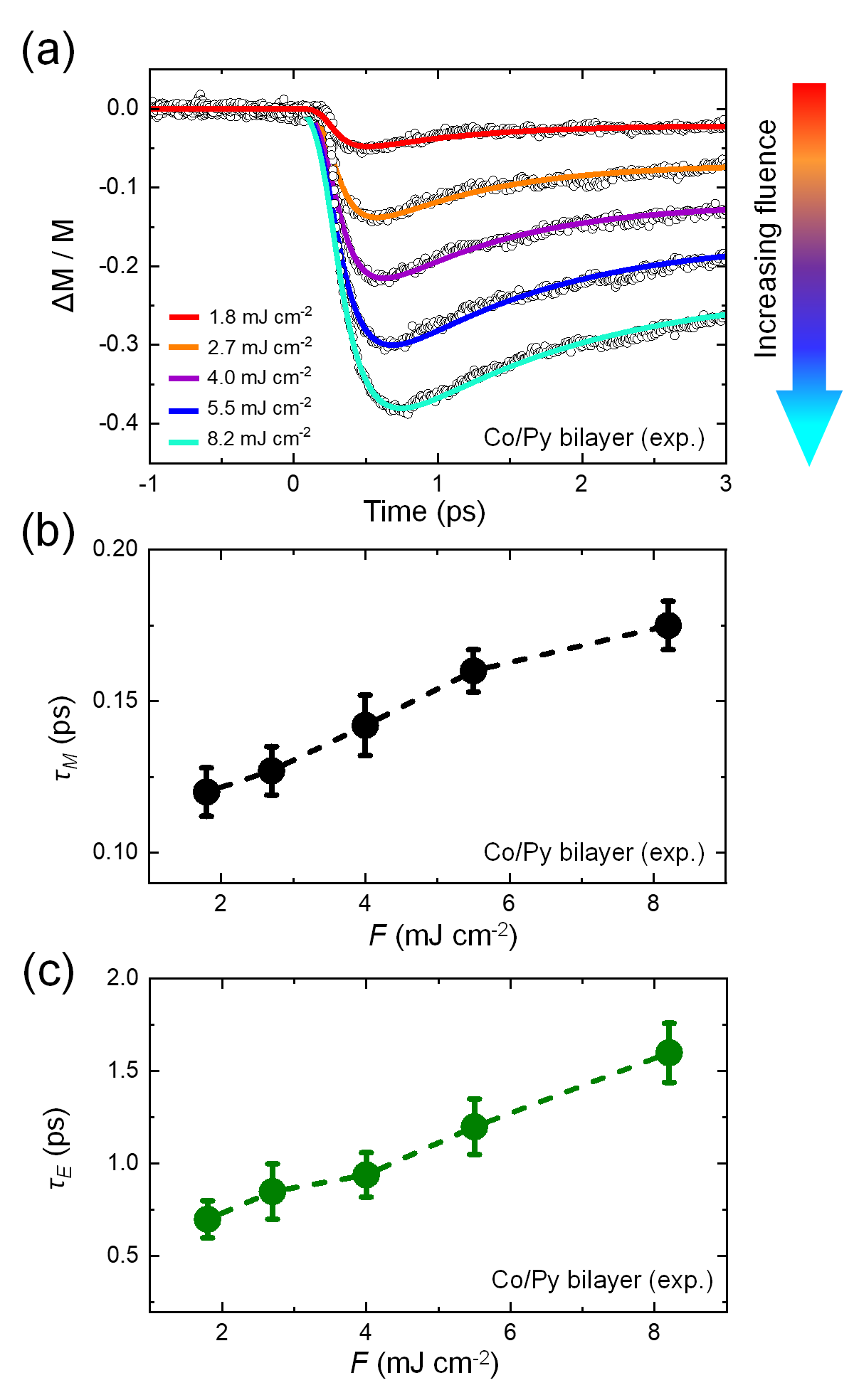}
	\caption{(a) Ultrafast demagnetization traces of the Co/Py bilayer at different pump fluences ($F$) fitted with the phenomenological three-temperature model (3TM). Black dots are experimental data and the solid lines of different colors are the fitting lines with the 3TM. Different colors indicate different $F$ values as mentioned in the graph. (b) Demagnetization times $\tau_M$ vs $F$, (c) fast remagnetization times $\tau_E$ vs $F$.}
	\label{fig3}
\end{figure}

We first study the ultrafast demagnetization and fast remagnetization processes (highly non-equilibrium regime after excitation by fs laser pulses) of the Co (\SI{1.5}{\nm})/Py (\SI{1.5}{\nm}) bilayer. We analyze the demagnetization dynamics using experimental methods and theoretical simulations.
\par
First, we compare the experimental demagnetization of the Co/Py  bilayer to the contributions of the individual Co and Py layers in a coupled bilayer using $\mu T$ model simulations. \autoref{fig:ultrafast_magnetization}(a) shows the experimental demagnetization traces of the Co/Py bilayer (solid line) and the theoretically calculated contributions of the Co (short-dashed-lines) and Py (long-dashed-lines) layers at two different excitation fluences. The demagnetization dynamics at lower and higher pump fluences are represented by red ($F$ = 1.8 mJ cm$^{-2}$) and blue ($F$ = 5.5 mJ cm$^{-2}$) lines, respectively. We perform high-resolution time-resolved Kerr rotation experiments for up to 10 ps from the zero delay with a temporal resolution of 30 fs. Our method probes the magnetization dynamics of the entire bilayer system.

A rapid decrease of magnetization takes place on the fs timescales, followed by a pronounced recovery. From the calculated demagnetization dynamics of Fig. \ref{fig:ultrafast_magnetization}(a), it is evident that at a particular excitation fluence, Co exhibits faster de- and remagnetization dynamics compared to Py. Such behavior has also been previously observed experimentally \cite{Anulekha2024PRB, Unikandanunni_APL_2021}. However, the experimental demagnetization traces of the Co/Py bilayer lie between those of the individual Co and Py layers. 
At early timescales, after ultrashort laser pulse excitation, the bilayer exhibits demagnetization dynamics that is faster than that of Py but similar to that of Co. 
However, the remagnetization dynamics of the bilayer is slower than that of Co and lies between Co and Py. 
Additionally, we can distinguish the contributions from the individual materials in the simulated remagnetization dynamics with a fast timescale in the beginning stemming from Co and a slower one later on stemming from Py.  

In Fig.~\ref{fig:ultrafast_magnetization}(b) we combine the numerically simulated demagnetization of the individual Co and Py layers to obtain the demagnetization dynamics of the entire bilayer. This is further compared to the experimentally measured data of the entire bilayer. We find good agreement with the experiment for a 3:1 superposition of layer-dependent contributions from the simulations. The simulated magnetization of the bilayer with the parameters used in Fig.~\ref{fig:ultrafast_magnetization}(a) is denoted as "weak coupling" in Fig.~\ref{fig:ultrafast_magnetization}(b) (long-dashed magenta line). We call the coupling weak because in the simulations in this case we find only small deviations from the magnetization of completely decoupled Co and Py layers [i.e., in the simulations we consider a bilayer formed of Co and Py layers, but without considering any coupling between the individual layers ("no coupling", solid green line)]. On the other hand, in the simulated case where the coupling between individual Co and Py layers is strong (for example, by increasing the interlayer coupling parameter by a factor of 500), the resulting magnetization deviates significantly from the experimental data ("strong coupling", short dashed brown line). We conclude that the experimental demagnetization of the bilayer is best described by weakly coupled or decoupled Co and Py layers within the bilayer.

\par
Next, we quantitatively compare the quenching vs. time required to reach the minima for the Co/Py bilayer, and individual Co and Py layers in Fig. \ref{fig:ultrafast_magnetization}(c). Here, it is evident that the time required to reach the minima increases steadily with increasing quenching, in line with previous results~\cite{Koopmans2010, Roth2012, suchetana2023, Anulekha2024PRB}. 
This behavior can be attributed to enhanced magnetic fluctuations, which result from a higher population of hot electrons at elevated electron temperatures, particularly when the electron temperature approaches the Curie temperature. Magnetic fluctuations can be linked to spin-flip processes caused by EY scattering, where an excited electron interacts with an impurity/phonon. This interaction alters the probability of the electron occupying a particular spin state, depending on the coupling strength between the electron and the spin system.

The observed behavior in Fig. \ref{fig:ultrafast_magnetization}(c) can also be explained in terms of chemical potentials. Quenching is determined by the transient equilibration between the spin-dependent chemical potentials following their non-equilibrium caused by the excitation pulse \cite{Mueller2014}. The higher the absorbed fluence, the further will the spin-dependent chemical potentials pushed away from each other, and the stronger will be the non-equilibrium that accompanies laser excitation. A stronger non-equilibrium implies longer time required to reach transient equilibration of the chemical potentials \cite{Mueller2014}. However, it should be noted that the slope of the increase is different for all three parts of the bilayer considered, as shown in Fig.~\ref{fig:ultrafast_magnetization} (c). The demagnetization is fastest in the Co layer and slowest in the Py layer, whereas, the experimentally measured total bilayer exhibits intermediate timescales.  The comparison between theory and experiment shows that on ultrafast timescales, during the reduction of magnetic order by the interaction with the ultrashort laser pulse, the dynamics of the individual materials in the bilayer are decoupled from each other. This implies that in the highly non-equilibrium regime (fs timescales), the intrinsic layer-dependent demagnetization processes has a larger influence on the dynamics than the potential interfacial coupling.
\par
We further analyze the demagnetization dynamics using an additional approach. We compare our experimental demagnetization traces with the phenomenological three-temperature model (3TM) \cite{DallaLonga2007}, which is obtained by solving the energy rate equation between three different degrees of freedom, e.g. electron, spin, and lattice (see the Supplemental Material for details). The TR-MOKE signals of the demagnetization evolution after ultrashort optical excitation measured at different excitation fluences ($F$) ranging from 1.8 mJ cm$^{-2}$  to 8.2 mJ cm$^{-2}$ are shown in Fig.~\ref{fig3}(a). The fit reveals an increase in both demagnetization time $\tau_M$ and fast remagnetization time $\tau_E$ with increasing fluence as shown in Fig.~\ref{fig3}(b) and (c), respectively. This behavior is in agreement with Fig.~\ref{fig:ultrafast_magnetization}(b). It is worth mentioning that the values of both $\tau_M$ and $\tau_E$ for the bilayer are lower than those obtained for the pure Py sample \cite{Anulekha2024PRB}.

\subsection{Dynamics in the close-to-equillibrium regime}

\begin{figure}
    \centering
	\includegraphics[width=3.0 in]{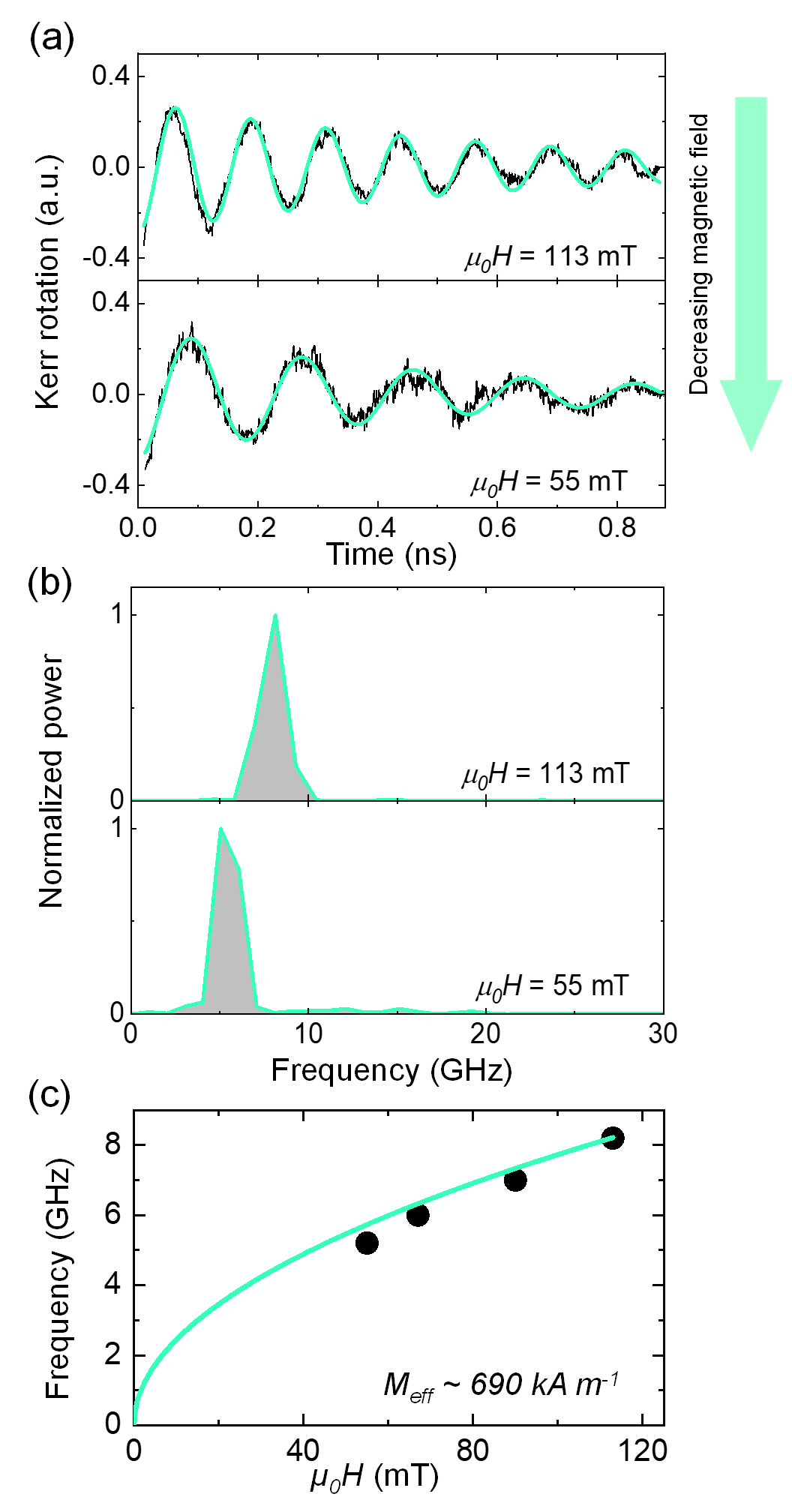}
	\caption{Magnetic field dependent precession dynamics: (a) Background-subtracted time-resolved Kerr rotation data for the Co/Py bilayer sample measured at three different magnetic fields and at $F$ = 8.2 mJ cm$^{-2}$. Brown Solid lines are fitting lines. (b) Corresponding FFT power spectra.  
    (c) Magnetic field dependence of precession frequency. Solid line represent the Kittel fit to the data points. 
    }
	\label{fig4}
\end{figure}

 We next study the precessional magnetization dynamics in the close-to-equillibrium regime well after the laser excitation. The precession dynamics in the gigahertz (GHz) range is governed by the phenomenological Landau-Lifshitz-Gilbert (LLG) equation \cite{gilbert2004}. Figure \ref{fig4} (a) shows the background-subtracted time-resolved precession data for two different magnetic fields, fitted with a damped sinusoidal equation,
\begin{equation}
	M(t) = M(0)e^{{-t}/{\tau_d}} \mathrm{sin}(2{\pi}ft)
	\label{eq3}
\end{equation}

where, $M(0)$ is the initial amplitude of the precession, $\tau_d$ is the precession relaxation time obtained as a fitting parameter, and $f$ is the precession frequency. The precession frequency can also be extracted directly from the fast Fourier transform (FFT) of the precessional oscillation. The FFT of the precessional oscillations yields a single magnon mode in each case (Fig.~\ref{fig4}(b)), indicating that on longer timescales, the individual layers of the bilayer precess together in a coupled way with the same frequency. The observation of coupled behavior is in contrast to some earlier reports \cite{Crew2005, Kennewell_JAP_2010, arabinda2014, Stenning2015}, where two modes were observed for bilayers with much higher thickness. Previously Crew et al.~\cite{Crew2005} reported two magnon modes in a thicker bilayer film Py (50 nm) / Co (100 nm) sample, which were termed optical and acoustic modes. Later Haldar et al.~\cite{arabinda2014} observed two modes in much thinner Py(10 nm - 30 nm)/Co(10 nm) bilayer system by Brillouin Light scattering technique, which were characterized differently as Damon-Eshbach (DE) and perpendicular standing spin wave (PSSW) modes. The contrasting observations of the magnon dynamics in the literature relative to our work is likely due to the bilayer studied here being much thinner as compared to those studied elsewhere. Since we have ultrathin layers, we can expect some high-frequency modes at $\gtrsim$ 100 GHz. However, we do not see any such modes. The effective magnetization $M_\mathrm{eff}$, which includes the saturation magnetization and potential additional out-of-plane (OOP) anisotropies, is calculated from the magnetic field dependence of the precession frequencies in Fig.~\ref{fig4}(c) and fitting the data points with the Kittel formula \cite{kittelformula},
\begin{equation}
	f = \frac{1}{2\pi}{\sqrt{\omega_\mathrm{H}(\omega_\mathrm{H}+\omega_\mathrm{M}}})
	  \label{eq4}
\end{equation}
where $\omega_\mathrm{H}=\gamma \mu_0 H$, $\omega_\mathrm{M}=\gamma\mu_0M_\mathrm{eff}$ and $H$ is the externally applied magnetic field and $\gamma = g\mu_B/{\displaystyle \hbar}$ is the gyromagnetic ratio with g = 2. From the fit, $M_\mathrm{eff}$ is obtained to be $\sim{690\pm 20}$ kA m$^{-1}$ for the Co/Py bilayer. This value is much smaller than Co ($\sim{1400}$ kA m$^{-1}$) and closer to that of Py ($\sim{760}$ kA m$^{-1}$) \cite{Anulekha2024PRB, suchetana2023}. 

\begin{figure*}
        \centering
	\includegraphics[width=6in]{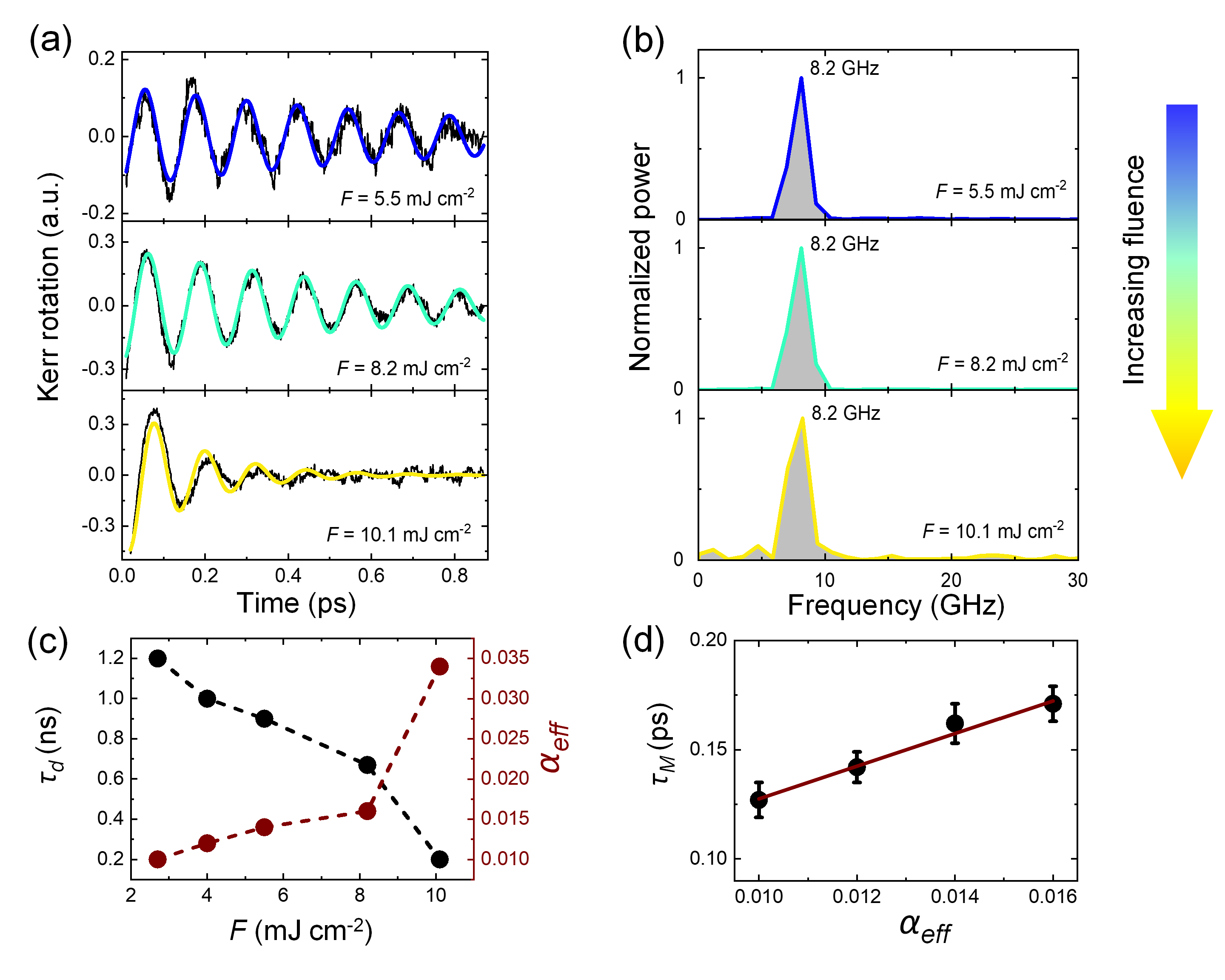}
	\caption{Pump fluence  dependent precession dynamics: (a) Background-subtracted time-resolved Kerr rotation data for the Co/Py bilayer sample measured at different pump fluences ($F$) and at magnetic fields and at $\mu_0H$ = 113 mT.  (b) Corresponding FFT power spectra. (c) Variation of precessional relaxation time $\tau_d$ and effective damping $\alpha_\mathrm{eff}$ with pump fluence (d) Direct correlation between demagnetization time and effective damping.}
	\label{fig5}
\end{figure*}

We next study the modulation of precessional dynamics with excitation fluence. Figure \ref{fig5}(a) shows the background-subtracted time-resolved Kerr rotation data of the bilayer measured at different pump fluences ($F$) and fixed magnetic field $\mu_0H$ = 113 mT. The corresponding FFT power spectra in Fig.~\ref{fig5}(b), shows a single precessional mode in each case which again indicates that on longer timescales the two layers precess together in a coupled way with the same frequency. It is worth mentioning that, in contrast to our earlier work on Ni \cite{akira2024jpcm}, here we do not observe any significant phase shift of the precessional dynamics with pump fluence. This is an indication of weak magneto-elastic (ME) coupling in this sample. We can calculate the effective damping factor $\alpha_\mathrm{eff}$ from the extracted value of precessional relaxation time $\tau_\mathrm{d}$ and $M_\mathrm{eff}$ as \cite{walowski2008, suchetana2023, Anulekha2024PRB}:

\begin{equation}
    {\alpha_{eff}}=\frac{1} {\tau_d \left( \omega_\mathrm{H}+\frac{\omega_\mathrm{M}}{2}\right)}
\end{equation}

Figure \ref{fig5}(c) shows an enhancement in $\alpha_\mathrm{eff}$ associated with a corresponding decrease in the precessional relaxation time $\tau_d$ as the pump fluence ($F$) increases. We  observe that $\alpha_\mathrm{eff}$ increases from 0.010 to 0.035 as $F$ increases from 2.7 to 10.1 mJ cm$^{-2}$. Various procedures for exciting long time scale (ns) precessional (magnon) dynamics show different mechanisms responsible for the exploration of different energy and angular momentum transfer channels. Microscopically, the EY spin-phonon interaction mechanism has been thought to be the main contribution to precessional damping \cite{Stiles2007, Fahnle2010, Sucheta2019}. Phenomenological models have suggested that the ratio between the temperature of the bilayer and Curie temperature transiently increases with increasing fluence leading to higher damping \cite{Nieves_PRB2014, Liu2016, Sucheta2019}.
However, the transient rise of system temperature at higher pump fluences in turn increases the probability of generation of more incoherent (or thermal) magnons relative to coherent magnons leading to dephasing driven by nonlinear magnon-magnon interactions at elevated temperature \cite{Anulekha2024PRB}.

\subsection{Bridge between highly non-equilibrium and close-to-equilibrium regime}

An important indicator of the dominant microscopic mechanism underlying ultrafast demagnetization is the correlation between demagnetization time and Gilbert damping \cite{Koopmans2005unifying, Fahnle2010, Fahnle2017PRB, suchetana2023}. Such a study can also facilitate a more comprehensive understanding of the relationship between highly-non-equilibrium (fs timescales) and close-to-equilibrium (ns timescales) regimes. In 2005, Koopmans et al. \cite{Koopmans2005unifying} theoretically calculated an inverse proportional relationship between demagnetization time ($\tau_M$) and damping based on quantum mechanical considerations. 
Nevertheless, this provoked vigorous debate as the validity of the model could not be confirmed experimentally when damping and demagnetization time were tuned by transition metal and rare earth doping \cite{walowski2008}. In 2010, Fähnle et al. demonstrated that the damping factor can be either directly or inversely proportional to $\tau_M$, contingent on the prevailing microscopic contribution to the magnetic damping \cite{Fahnle2010}. Later, a direct proportional relation between the two parameters in Co/Ni bilayers \cite{Fahnle2017PRB} and individual Co, Ni, and Permalloy thin films \cite{suchetana2023} were experimentally demonstrated.  To study the correlation between ultrafast demagnetization time ($\tau_M$) and effective damping ($\alpha_\mathrm{eff}$) in a system, we have used the excitation fluence to systematically modulate both $\tau_M$ and $\alpha_\mathrm{eff}$. Figure \ref{fig5}(d) shows a direct proportionality between $\tau_M$ and $\alpha_\mathrm{eff}$ in the Co/Py bilayer, which can be understood phenomenologically by their individual relation to the absorbed fluence as well as microscopically through their intrinsic correlation via spin-dependent electron-hole pair generation.

The observed behavior in Fig.~\ref{fig5}(d) can be phenomenologically explained as follows. The demagnetization time $\tau_M$ increases with increasing fluence as shown in Fig. \ref{fig3}(b). This is a general phenomena and has also been reported earlier \cite{Koopmans2010, Atxitia_2010, Roth2012, suchetana2023, Anulekha2024PRB}. This increment with fluence can be explained using the enhanced magnetic fluctuations (creation of a larger population of hot electrons) at elevated temperatures, when the electron temperature is close to the Curie temperature. The magnetic fluctuations can be related to the spin flip processes due to the EY scattering of an excited electron with an impurity/phonon, which changes the probability of finding the electron in one of the spin states, which depends on the coupling strength between the electron and the spin system. Furthermore, a higher fluence leads to separation of the spin-dependent chemical potentials further apart, consequently resulting in increased demagnetization times. On the other hand, looking into the long time scale (ns) magnon dynamics, we observe that the effective damping $\alpha_\mathrm{eff}$ also increases with the absorbed fluence (Fig. \ref{fig5}(c)) as a result of increased non-linearity in the magnon system due to the transient rise in system temperature leading to the generation of more incoherent magnons at higher fluences \cite{Anulekha2024PRB}. 

\par
The behavior of Fig.~\ref{fig5}(d) can also be understood from a microscopic perspective. As demonstrated by Gilmore et al. \cite{Stiles2007}, Gilbert damping in metallic ferromagnets is primarily attributed to the precession (magnon) dynamics generating excited electron-hole ($e-h$) pairs, which subsequently undergo spin-dependent scattering within the lattice, resulting in the transfer of energy and angular momentum from the electronic system to the lattice. The generation of $e-h$ pair in the same band (intraband mechanism), leads to a conductivity-like contribution to damping, and the generation of $e-h$ pair in different bands (interband mechanism), leads to a resistivity-like contribution to damping. Accordingly, Fähnle et al. derived a model based on the breathing Fermi surface model \cite{Fahnle2011JPCM, Fahnle2010}. 
It assumes that the intraband pairs are generated due to the changing spin-orbit interaction as the orientation of the homogeneous magnetization changes over time. Consequently, some states that are just below the Fermi surface for one orientation are pushed above the Fermi surface, whereas other states that were originally above are pushed below, which means that the Fermi surface “breathes” during the precessional dynamics. 
Subsequently, the $e-h$ pairs generated by precession relax through EY scattering processes with the lattice, thereby dissipating their energy and angular momentum to the lattice, the strength of which depends on how far from equilibrium the system gets; thus, affecting the damping. On the other hand, the observed behavior of the laser-induced ultrafast demagnetization can be microscopically explained by EY type spin-flip-scattering processes, according to which the excited electrons after fs laser pulse excitation flip their spin when scattering with impurities or phonons. 
As for both demagnetization and damping, the angular momentum of the spin has to be transferred from the electronic spin system to the lattice, we can consider a similar type of spin-flip scattering mechanism to be relevant for the two situations, and hence, the observed direct proportional relation in Fig.~\ref{fig5}(d) can be microscopically understood in terms of EY processes on both timescales.

\section{Summary and Conclusion}
In summary, we have studied the magnetization dynamics of an ultrathin Co(1.5 nm)/Py (1.5 nm) bilayer from femtosecond (fs) to nanosecond (ns) timescales. We observe that the interfacial effects in such bilayer is dynamic (i.e., timescale dependent). On fs timescales, the dynamics indicate that the constituent layers of the bilayer behave decoupled from each other when the system is in a highly non-equilibrium regime caused by ultrashort laser excitation. However, on longer (ns) timescales, when the system is in close to equilibrium, the individual layers of the bilayer precess together as an effective single layer in a coupled manner with a single frequency. Furthermore, a proportional relationship between the demagnetization time and effective damping in the bilayer indicates that the localized Elliott-Yafet (EY) spin-flip scattering is the dominant mechanism for ultrafast magnetization dynamics in the system. Since both demagnetization time and damping require a transfer of angular momentum from the electronic system to the lattice, the unification of these two seemingly unrelated parameters can facilitate the exploration of the microscopic mechanism of laser-induced magnetization dynamics from femtosecond to nanosecond timescales. Such a detailed analysis of different non-equilibrium regimes  of magnetization dynamics and the attempt to find a bridge between them in an exchange coupled, epitaxially grown Co/Py ultrathin bilayer will expedite its use in technological applications.

\section*{Author Contribution}
A.D. and C.S. contributed equally. A.D., P.H., and A.L. designed and performed TR-MOKE experiments. A.D. analyzed the experimental results in consultation with M.A., G.v.F., and P.P., and L.S. fabricated the sample. C.S. performed $\mu T$-model simulations in consultation with B.R., and S.A. All authors contributed to the analysis and discussion for the results. A.D., C.S., and S.A. wrote the first draft of the manuscript. All authors contributed to the editing of the manuscript.
\\
\section*{Acknowledgements}
The authors are thankful to B. Hillebrands, H. C. Schneider and B. Stadtm{\"u}ller for helpful discussion. The authors appreciate the Allianz für Hochleistungsrechnen Rheinland-Pfalz for providing computing resources on the Elwetritsch high-performance computing cluster. The authors gratefully acknowledge financial support of the Deutsche Forschungsgemeinschaft (DFG, German Research Foundation) through the SFB/TRR-173-268565370 “Spin+X” (projects A08, B03 and B11).

\bibliography{all.bib}

\end{document}